# Re-orienting the Turbulent flow over an Inclined Cylinder of Finite Aspect ratio

Harshad Kalyankar[1], Lutz Taubert[2], and Israel Wygnanski[3]
*Aerospace and Mechanical Engineering, University of Arizona, Tucson, AZ-85721*

**Turbulent flow around a swept back circular cylinder was investigated experimentally using one or two small sweeping jets emanating tangentially to the surface and orthogonally to the axis of the cylinder creating a yawing moment that overcomes the natural restoring force to the plane of symmetry. It appears from the integral force balance data that large yawing moment coefficients can be generated in this manner, and the results could be used in the orientation and attitude control of a refueling boom thus avoiding the 'H' shape control surfaces that are currently used on Air Force tankers. The interaction between the jet and the flow in the lee of the cylinder was mapped using surface oil flow visualization and 2D-Particle Image Velocimetry technique. The sensitivity of the interaction and its outcome to the change in the azimuthal location of the actuators was investigated.**

## Nomenclature

| | | |
|---|---|---|
| $AR$ | = | aspect ratio of the cylinder assembly $\approx 9.1$ |
| $d$ | = | cylinder diameter = 0.0762 m |
| $L$ | = | length of cylinder assembly = 0.6934 m |
| $c$ | = | freestream chord length = $d/\cos(\Lambda)$ m |
| $A_{cylinder}$ | = | sectional area of cylinder = $L*d \approx 0.0528$ m$^2$ |
| $A_{tip}$ | = | sectional area of elliptical tip $\approx 0.0059$ m$^2$ |
| $A_{total}$ | = | total sectional area of cylinder assembly $(A_{cylinder} + A_{tip}) \approx 0.0587$ m$^2$ |
| $U_\infty$ | = | freestream velocity (m/s) |
| $q_\infty$ | = | freestream dynamic pressure (N/m$^2$) |
| $Tu$ | = | turbulence intensity (%) |
| $SF$ | = | side force (N) |
| $YM$ | = | yaw moment (Nm) |
| $D$ | = | drag force (N) |
| $C_Y$ | = | side force coefficient = $\dfrac{SF}{q_\infty A_{total}}$ |
| $C_{LN}$ | = | yaw moment coefficient = $\dfrac{YM}{q_\infty A_{total} c}$ |
| $C_D$ | = | drag coefficient = $\dfrac{D}{q_\infty A_{total}}$ |
| $\Delta C_{LN}$ | = | difference between the actuated and baseline case of $C_{LN}$ |
| $C_\mu$ | = | momentum coefficient |
| $Re$ | = | reynolds number = $\dfrac{U_{inf} D/\cos(\Lambda)}{\nu}$ |
| $P$ | = | supply air pressure for the sweeping jet actuators |
| $y_w$ | = | wake half-width, $w$= +1/2 for top half-width, and $w$= -1/2 for bottom half-width |

**Greeks**

| | | |
|---|---|---|
| $\Lambda$ | = | sweep-back angle (°) |
| $\beta$ | = | yaw angle (°) |
| $\zeta_x$ | = | azimuthal angle of feature 'X' from the front attachment line at $\beta$=0° (°) |
| $\zeta_o$ | = | azimuthal angle of the upstream lip of the actuator slot (°) |
| $\zeta_{sep}$ | = | azimuthal angle of the separation line (°) |

---

[1] Graduate Research Assistant, AIAA Student Member
[2] Research Assistant Professor, AIAA Member
[3] Professor, AIAA Fellow, Corresponding Author, wygy@email.arizona.edu

## I. Introduction

VORTEX shedding characterizes the flow around bluff bodies, and is the most ubiquitous feature exploited in some applications in the aerospace and civil industries, while plaguing others. Wings buffet and flutter, suspended cables gallop and might have wind induced vibration [1, 2], chimneys, cooling towers, and even suspension bridges may collapse if they happen to resonante with the vortex shedding frequency. The flow over inclined cylinders of 'infinite' span has its applications for electrical transmission lines (high aspect ratio under oblique wind) and marine structures under sheared currents [3]. These flows have three-dimenional effects and are of fundamental interest in engineering, because of the nature of the stagnation line, leading to attachment line instabilities [4, 5, 6], its transition to turbulence, and vortex shedding. Wake flows for inclined cylinders (infinite aspect ratio, simulated by endplates) were studied experimentally by Thakur et. al. [7], where the applicability of quasi-two dimensional assumption was assesed for inclination angles of 0º, 30º, and 60º. The wake vortex orientation in the volumeteric Laser Induced Fluorescence (LIF) was observed to be parallel to the cylinder axis for $\Lambda = 0º$ and 30º, but was at a lower inclination angle for 60º case. These results were reinforced by Zhou et al. [8], where the dependence of wake on $\Lambda = 0º$ to 45º was investigated incorporating multi-hot wire probes over a streamwise range of $x/d = 10$-$40$. The Independence Principle or the Cosine Law, defined as the independence of the flow parameters, here Strouhal number ($St_N/St_0$), was observed for $\Lambda = 0º$ to $\Lambda \approx 40º$ (within experimental uncertainty). $St_N$ was calculated considering normal velocity component (considered here to be the azimuth direction). The inclined orientation of the wake vortex with the cylinder axis at $\Lambda = 60º$ for Thakur et. al. [7] confirms the departure from Independence Principle at $\Lambda > 40º$. Marshall [9] performed pertubation analysis to expand the applicability of quasi-two dimensional approach to finite length cylinders, limiting it to positions $L$ along the cylinder where the criterion $\frac{d}{L} \ll \frac{Normal\ U_\infty\ component}{Axial\ U_\infty\ component}$ was satisfied, thus requiring that only cross-stream $Re$ (considered here as azimuth direction) be included in the governing equations. Marshall also noticed that at large values of $\Lambda$, the cross-stream vortex street (spanwise vorticity) in the cylinder near-wake and the axial flow deficit in the downstream vortex cores lead to an instability in the vortex sheet, thus breaking down the Independence Principle.

A plethora of applications dealing with maneuvering aircraft, submarines, torpedoes, missiles, and refuelling booms [10] can be represented by dynamically moving inclined cylinders of finite *AR* with swept back or forward configurations. When the inclined cylinder is also yawed all symmetries are broken and the level of complexity increases [11]. Limiting the infinite length consideration, Ramberg [12] investigated the end effects of a yawed cylinder on its vortex wake properties, including the shedding frequency and angle, vortex-formation region length, and the wake width, with the change in cylinder length, and found the results to be sensitive to the end conditions specially at low *Re*.

Controlling the crossflow over the circular cylinder poses a challenge because the separation line is not geometrically determined. At sub-critical Reynolds numbers, Naim et al. [13] observed that separation can be controlled by two distinctly different mechanisms, namely: by forcing laminar-turbulent transition when applied at $30º < \zeta < 60º$ from the forward stagnation point; and by directly forcing the separated shear-layer at $\zeta > 90º$. They further observed that the universal Strouhal law also holds for active control on cylinders, as long as the excitation frequency is significantly higher than the natural vortex-shedding frequency. This extension of the universal Strouhal law is attributed to separation control, which narrows the wake, thereby increasing the vortex shedding frequency and decreasing the drag.

The current study focuses on understanding the flow over a finite *AR* cylinder (swept-back at a fixed $\Lambda$), with variable yaw angles; in order to identify avenues where AFC could be applied and investigate its influence. Sweeping jet (SWJ) actuators were employed in the current study. The effect of surface curvature on the development of two-dimensional turbulent wall jet was experimentally investigated by Neuendorf and Wygnanski [14]. The maximum velocities ($U_{max}$) in the curved wall jet was observed to decrease faster with streamwise travel (*x*) than the approximate $U_{max} \propto x^{-1/2}$ relation for a plane wall jet, suggesting an increased rate of spread due to the entrainment by large eddies. It was observed that this tangential deceleration lead to a generation of a mean velocity component normal to the surface; the wall jet flow separated when this normal velocity component became comparable with the tangential one.

## II. Experimental Setup

**Wind Tunnel Facility:**

The experiments were carried out in the Aerospace and Mechanical Engineering subsonic Research Wind Tunnel at the University of Arizona. The wind tunnel has a 0.9 m x 1.2 m x 3.65 m (3 ft x 4 ft x 12 ft, height x width x length) test-section and is a closed-loop facility (fig. 1) capable of $U_\infty$=80 m/s (262.5 ft/s). The air temperature was controlled to be 22ºC (72ºF) within ±0.5ºC (1ºF) through the use of a heat exchanger and chilled water supply. The turbulence intensity is $Tu \leq 0.15\%$ for $f_{lowpass}$ = 10 kHz and $Tu \leq 0.05\%$ for $f_{bandpass}$ =1 Hz–10 kHz.

**Swept-back cylinder model:**

The cylinder model of 3" (0.0762 m) diameter and 24" (0.6096 m) span was mounted on a five-component side-wall force balance via an adapter with $\Lambda$=60º sweep-back (fig. 2), resulting in $AR \approx 9.1$. A 14" diameter circular endplate near the tunnel wall was used to separate the cylinder from the side-wall tunnel boundary-layer. The coordinate system used for forces and moments is defined in fig. 2. The moment center was selected to be the point where the cylinder axis meets the 14" diameter circular endplate. A tangential slot was machined parallel to the cylinder axis along the span for actuation purposes. The entire cylinder except the actuator exit areas was covered in a black foil, sealing the slot and providing optimal surface conditions for the used oil flow visualization. Although filled with model clay and covered, this slot, the joint of the two halves of the cylinder opposite the slot, the overlap of the foil, screw holes etc. introduced deviations from the ideally symmetric shape of the model. In the configuration shown in fig. 2, the cylinder is positioned at a negative yaw angle $\beta$. Figure 3a defines the azimuthal angle ($\zeta$) for any feature, from the front attachment line at $\beta$=0º, with the location of the slot indicated as $\zeta_o$, and the separation location as $\zeta_{sep}$.

**Active Flow Control (Sweeping Jet Actuators):**

3D printed sweeping jet actuators with a throat area of 0.050" x 0.050" and an outlet area of 0.050" x 0.5" were utilized. Two actuators were located at $3.8d$ and $6.47d$ (~1/3$^{rd}$ and ~2/3$^{rd}$ span) in the tangential slot. They were inclined and recessed in the slot (fig. 3b), using slot curvature and coanda effect to blow out tangentially to the cylinder's surface.

**Surface Oil flow Visualization technique:**

A surface oil flow visualization technique was utilized to qualitatively study the boundary-layer flow. This technique used a mixture of kerosene, fluorescing aviation oil and fumed silica particles. The mixture in the ratio of 2:1 (kerosene: oil) by weight with 1% weight of silica particles (both hydrophilic and hydrophobic), was mixed and applied to the black foil wrapped around the cylinder. With the boundary layer flow, the mixture moved due to local skin friction and revealed the surface flow topology. The cylinder was then illuminated under UV lights, causing fluorescence of the aviation oil. Three cameras set up fix on tripods were used to take images of the front, top and lee views. UV filters were used to block the illuminating UV light and allow only light emitted by the fluorescing streaklines to pass, resulting in high-contrast flow visualizations.

**Particle Image Velocimetry (PIV) technique:**

A LaVISION 2D-Particle Image Velocimetry (PIV) system was employed to investigate the effect of AFC on the near-wake (< 2$d$) azimuthally downstream of the cylinder at ~74.5% span (considering the span-length of the entire assembly). The camera axis had to be inclined (~20º) to the cylinder axis (fig. 4) due to limited optical access. The Nd-YAG laser sheet was aligned perpendicular to the cylinder axis. The error in the +Z-direction flow velocity due to the inclined camera axis was corrected during post-processing of the images. Due to such an inclination, perspective view of the cylinder was captured by the camera, hence the flow topology could not be captured at the bottom-half of the cylinder. A Nikon AF-S Micro-Nikkor 105mm focal length lens was utilized to capture the movement of tracer particles with a time delay of 18µs, resulting in a maximum particle displacement of 8 pixels. The standard DaVIS multi-pass image correlation analysis was performed, starting with an interrogation window of 64x64 pixels and refining to 32x32 pixels.

American Institute of Aeronautics and Astronautics

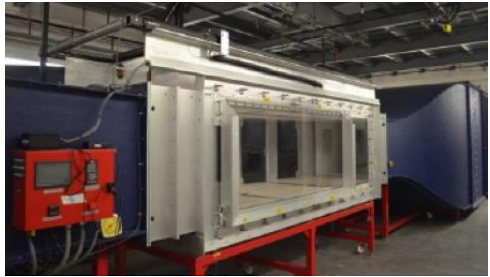
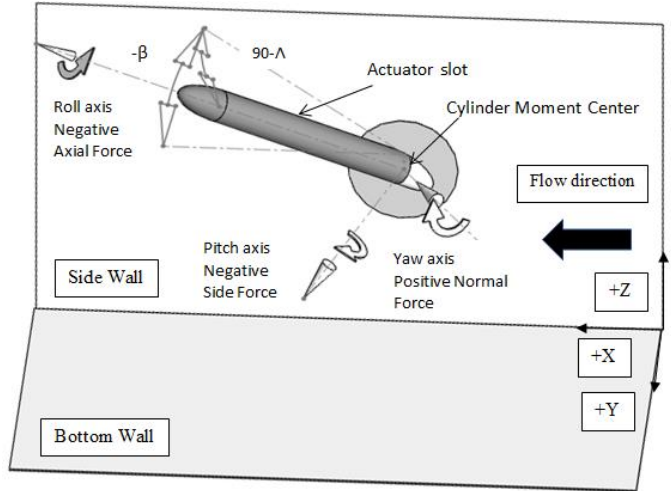

Figure 1. University of Arizona's 3' x 4' x 12' Wind tunnel test section

Figure 2. Sketch of swept-back Cylinder setup

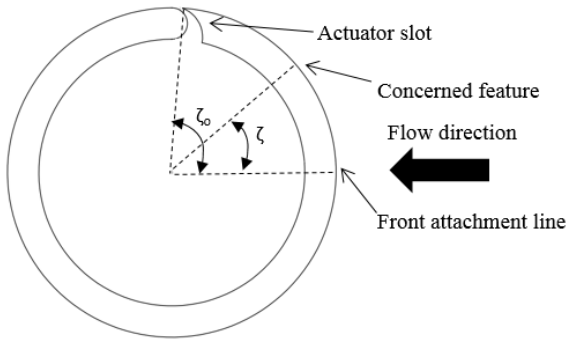
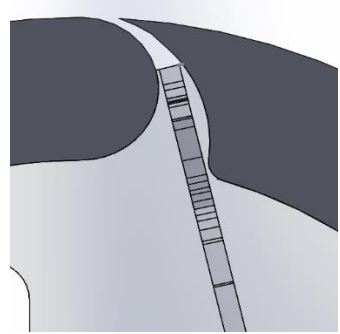

Figure 3a. Definition of azimuthal angle ($\zeta$)

Figure 3b. Placement of inclined actuator

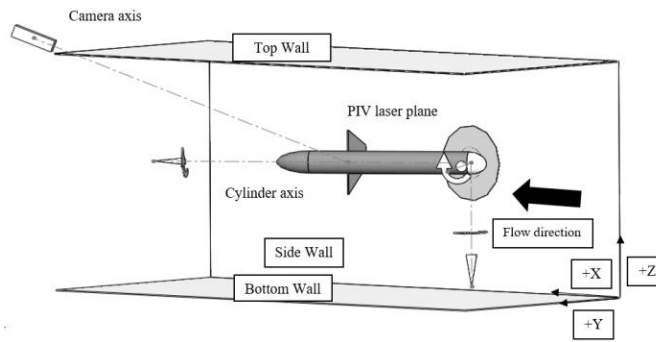

Figure 4. Inclined 2D-PIV setup

## III. Baseline Results

**1) Reynolds number variation and attachment line instability:**

    The dependence of drag on the traditional Reynolds number (based on $U_\infty$ and $d/cos(\Lambda)$) was determined for the swept back cylinder of $AR= 9.1$, at sweep back angle $\Lambda= 60°$ in the absence of yaw ($\beta= 0°$). The freestream velocity was increased from 5 m/s to 35 m/s at increments of 5 m/s resulting in $0.7*10^5 < Re < 3.3*10^5$. In the absence of trip-strip the $C_D$ reached a maximum value of ~0.25 at $Re \approx 1.5*10^5$, thereafter falling to $C_D \approx 0.19$ at the highest $Re$ tested (fig. 5). Placing a hot wire near the leading edge stagnation line (it is an attachment line since the spanwise component of velocity does not stagnate) close to the tip of the cylinder (just inboard of the rounded nose) revealed the existence of turbulence at the lowest free stream velocity tested $U_\infty= 2$ m/s and the transition location moved inboard with

American Institute of Aeronautics and Astronautics

increasing $U_\infty$. These observations were attributed to attachment line instability. The current experimental set up was not suited for this investigation because the thickness of the upstream boundary layer to the cylinder assembly was alleviated using a 14 inch diameter circular endplate, but could not be entirely eliminated; and it created a small necklace vortex at the root of the cylinder. Consequently transition to turbulence due to this instability was considered beyond the scope of the current investigation. Applying a circumferential boundary layer trip-strip at the root of the cylinder located at $0.8d$ from the plate while retaining two spanwise trip strips at $\zeta=\pm 30°$, fixed the spanwise transition location making it $Re$ independent at $Re > 1.9*10^5$, with $C_D \approx 0.2$ (fig. 5).

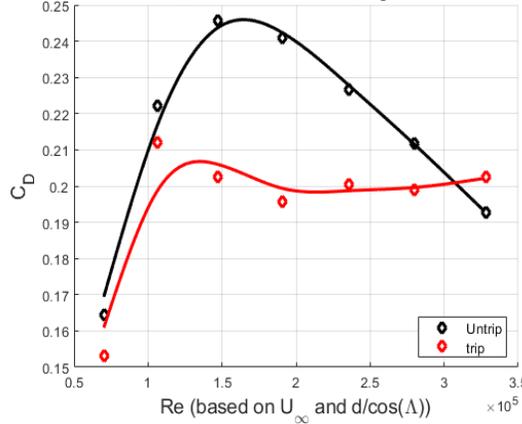

**Figure 5. $C_D$ variation with Re for trip and untrip case with $\beta=0°$ and $\zeta_o=90°$**

**2) Force balance data and Oil flow images:**

It is imperative to understand the flow field of a swept-back cylinder, since it is not as widely studied as the forebody problem which considers a swept forward cylinder with a slender forebody (a rounded or sharp nose cone) at moderate sweep (incidence) angles. Only then one may explore the various approaches for controlling the flow, altering the forces and moments acting on the body. Figures 6 and 7, display the integral values of side force ($C_Y$) and the restoring yawing moment ($C_{LN}$) as a function of the yaw angle ($\beta$) at $Re \approx 3.7*10^5$. For a 'perfectly symmetric cylinder', $C_Y = C_{LN} = 0$ as long as $\beta=0°$. But, the presence of the taped over actuator slot at azimuthal angle $\zeta_o$ resulted in flow asymmetry as shown in fig. 6, with the trip strips relocated and maintained at $\zeta=\pm 30°$. The worst asymmetry occurred at $\zeta_o=90°$ where the azimuthal velocity component was the highest and thus the effect of the tape thickness and jet nozzles on the side force was most notable. The restoring yawing moment for various values of $\beta$ is shown in fig. 7. In this case as well, the largest yawing moment at $\beta=0°$ corresponds to the taped slot location of $\zeta_o=90°$. Since the circular disc at the base of the cylinder was metric (i.e. it was attached to the cylinder and hence the balance) its drag was included in the force budget. One could subtract the effect of the disc assembly alone but the influence it has on the oncoming freestream flow, and the endplate-cylinder interaction could not be isolated, leading to the non-linear variation of $C_{LN}$, observed in fig. 7.

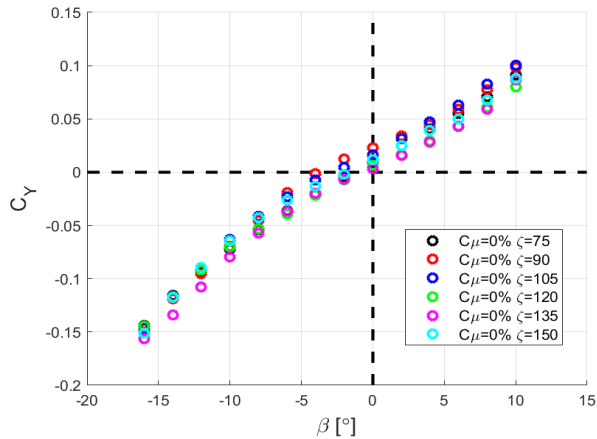 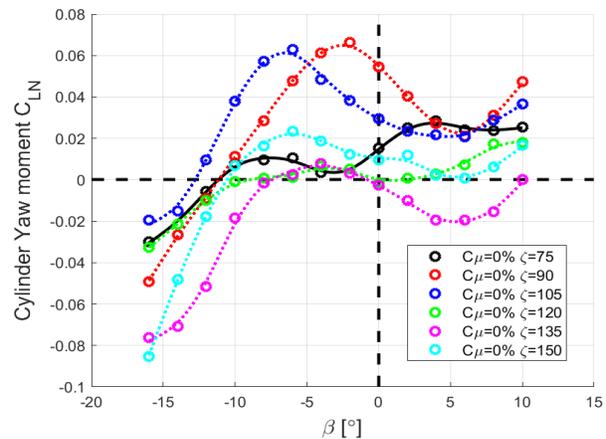

**Figure 6. $C_Y$ Vs. $\beta$ for various $\zeta_o$ locations**   **Figure 7. $C_{LN}$ Vs. $\beta$ for various $\zeta_o$ locations**

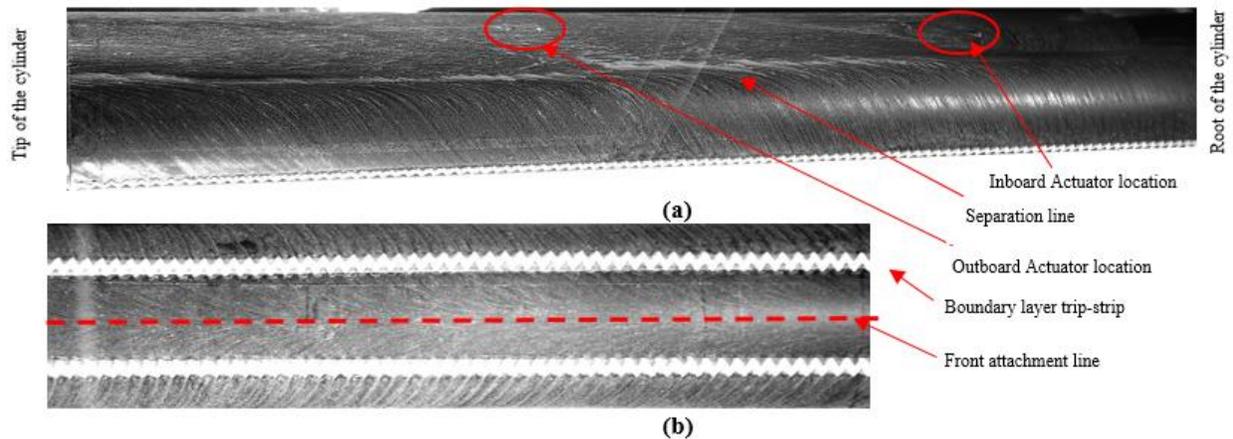

**Figure 8. Surface Oil flow visualization for $C_\mu$= 0%, $\beta$=0° and $\zeta_o$ =150°. a) Top view b) Front view of the cylinder.**

Oil flow over the swept back cylinder at $\beta$= 0°, with the slot containing the two sweeping jet actuators located at $\zeta_o$=150° is shown in Fig. 8. The front attachment line (fig. 8b) is at $\zeta$= 0° proving symmetry of the flow around the cylinder when the slot is downstream of the separation line (fig 8a), which is characterized by the accumulation of oil and silica particles at $\zeta \approx$ 115°. There is of course another separation line at $\zeta \approx$ -115° which is out of sight. In this case both $C_Y$= $C_{LN}$= 0 (figs. 6 & 7) as indeed they should be for perfectly symmetrical flow around the cylinder. The outboard flow near the front attachment line changes its direction due to the azimuthal acceleration that follows with increasing $\zeta$. The zig-zag trip strip generates streamwise vortices that are good indicators of the flow direction. As the azimuthal flow decelerates on the lee side of the cylinder, the surface streamlines turn outboard prior to separation. At $\zeta > \zeta_{sep}$ the surface flow is mostly directed along the span, but seems to be chaotic. For $\beta$= 0°, both attachment and separation lines are parallel to the axis of the cylinder. The flowfield on the inboard side of the lee-region was affected by the necklace vortex but it was impossible to quantify this effect.

For $\beta \neq 0°$, the front attachment line moves either upstream or downstream depending on the sign of $\beta$, and still be parallel to the cylinder's axis (fig. 9). The separated flow in the lee of the cylinder is no longer parallel to the axis of the cylinder and its inclination depends on $\beta$ (fig. 10) When $\beta < 0°$ the separation line is inclined towards the leading edge (i.e. downward in fig. 10a) while its inclination angle has an upward trend (toward the trailing edge) when $\beta>0°$. One may note that the outboard actuator located at $\zeta_o$=120°, highlighted in figure 10 is in the attached flow region when $\beta$ = -10° and is in the separated region when $\beta$ = +10°.

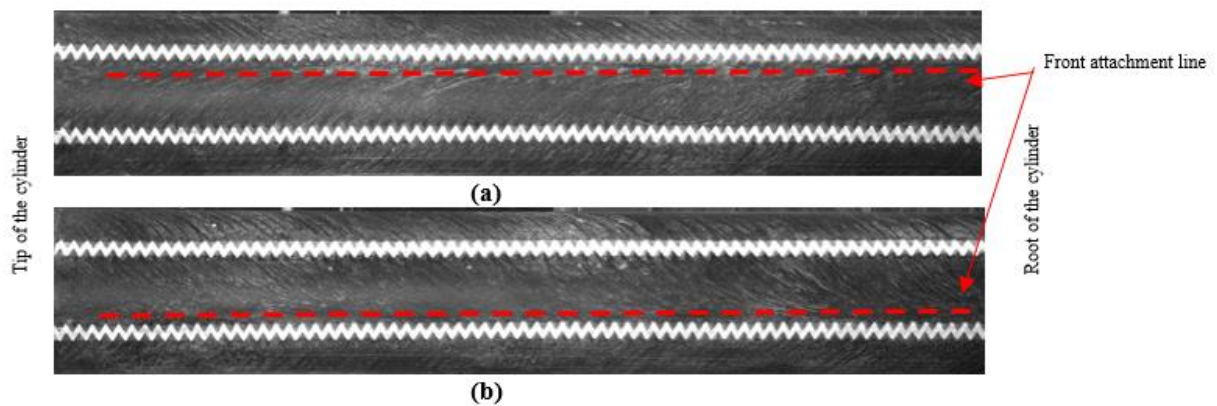

**Figure 9. Surface Oil flow visualization (Front view) for $C_\mu$=0% and $\zeta_o$=150°. a) $\beta$ = -10°  b) $\beta$ = +10°**

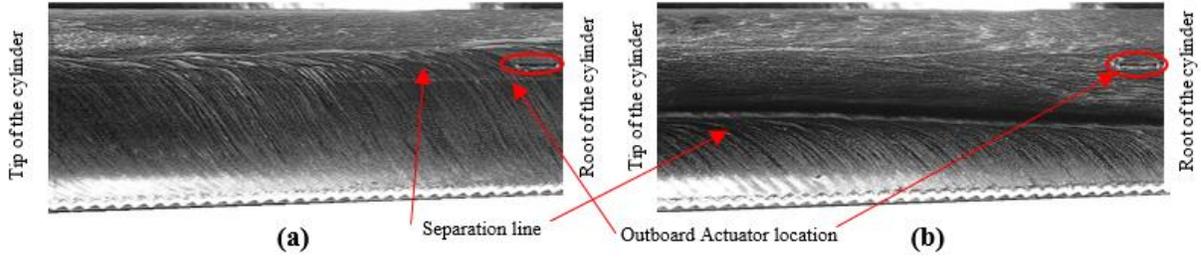

**Figure 10. Surface Oil flow visualization (Top view) for $C_\mu$=0%, $\zeta_o$=120; a) $\beta$= -10° b) $\beta$= +10°**

## IV. Active Flow Control (Sweeping Jet Actuators)

The flow over the swept back cylinder could be controlled by various approaches, the one utilized here were two sweeping jet actuators located in a thin spanwise slot that was entirely filled with clay and taped over except at the location of the actuators themselves. The two actuators divided the span into three equal segments and their location was not altered in the present report. Since the cylinder could be rotated around its axis the azimuthal location of the actuators $\zeta_o$ could be altered at will, as was the aggregate momentum output of the two sweeping jets, $C_\mu$. The yaw angle (i.e. the side slip angle, $\beta$) could also be changed. Finally, the performance of individual actuators in the $\zeta$ and $\beta$ domain was explored to understand the significance of their location and the level of their interaction when both are actuated.

### A. $C_\mu$ variation:

The effect of the two sweeping jets on the yawing moment of the cylinder relative to its moment center is shown in fig. 11. The actuator supply pressure was varied from 10-60 psig in increments of 10 psi, thus varying the flowrate through the actuators and their output $C_\mu$ values. The variation of $C_{LN}$ with $\beta$ for 0.2% ≤ $C_\mu$ ≤ 2.73% ($P$=10-60 psi), is compared with the baseline configuration when $\zeta_o$= 90° in fig. 11. A considerable yawing moment was generated by the asymmetric actuation. By plotting $\Delta C_{LN}$ representing the difference between the actuated ($C_\mu \neq 0$) $C_{LN}$ and the baseline ($C_\mu = 0$), versus $C_\mu$ for $\beta = 0°$ at $\zeta_o = 90°$, one realizes that $\Delta C_{LN}$ is linearly proportional to $\sqrt{C_\mu}$ (fig. 12) suggesting that jet entrainment may be the leading mechanism for generating side force. This was investigated by Wygnanski [15] and Wygnanski and Newman [16], and it was observed that the turbulent jet issuing normally, could be replaced by a line sink of variable strength that facilitates entrainment of the surrounding fluid. This entrainment increases the camber of the profile, and reduces or even eliminates the adverse pressure gradient upstream of the slot.

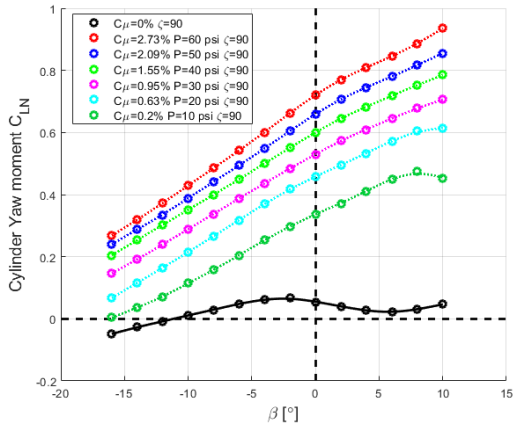

**Figure 11. $C_{LN}$ Vs. $\beta$ for various $\zeta_o$ locations**

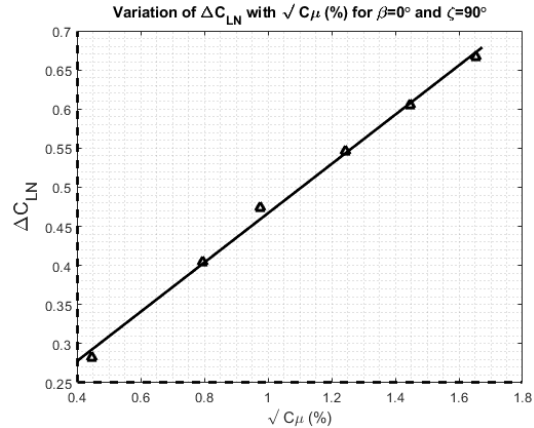

**Figure 12. $\Delta C_{LN}$ Vs. $C_\mu$ for $\zeta_o$= 90° and $\beta$= 0°, investigating $C_\mu$ variation**

In the absence of actuation the yawing moment (that is equivalent to pitch on a wing) $C_{LN} \approx 0$, vanishes when cylinder is aligned with the flow. Actuation from one or two of the sweeping jets installed changes that, and can balance the restoring moment at $\beta \neq 0°$. This restoring moment generated by AFC was dependent on the azimuthal location of the actuators (i.e. the slot) $\zeta_o$, on the yaw angle $\beta$, and on the momentum input $C_\mu$. The variation of $\Delta C_{LN}$ with these variables for a prescribed value of $C_\mu= 2.73\%$ emanating from two actuators is shown in fig. 13, indicating a strong coupling between $\beta$ and $\zeta_o$. In the absence of actuation at $\beta = 0°$ the flow separates naturally around $\zeta \approx 115°$ enclosing a vortex flow that is similar to a LE vortex on a swept back wing (see the schematic inset in fig. 13), as was confirmed by PIV. Consequently, when an actuator was located at $\zeta_o=120°$ (i.e. very close to the natural separation line), its effect was large but if it was moved to $\zeta_o=150°$ its effect was minimal. Surface oil flow visualization confirmed and assisted in explaining the balance measurements because the separation line was clearly visible, shown in fig. 14, where streaklines originating from various spanwise locations coalesce into a single line at $\zeta_{sep}$. The side force generated by AFC can be explained by the movement of the separation line farther downstream in the azimuthal direction. When AFC was effective (i.e. at $\zeta_o=120°$, fig. 14 center) the separation line moved beyond the actuator nozzle (encircled in red). Actuation from $\zeta_o=150°$ did not displace the separation line from its original $\zeta_{sep}$ (fig. 14 right).

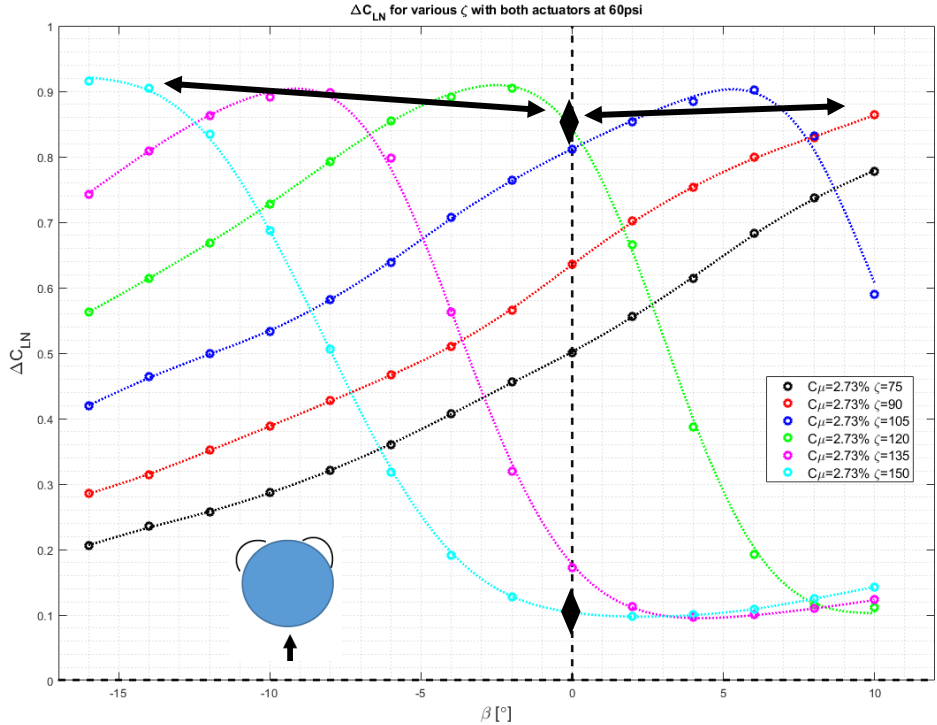

**Figure 13. Dependence of $\Delta C_{LN}$ on $\beta$ for $C_\mu= 2.73\%$ at various $\zeta_o$ locations.**

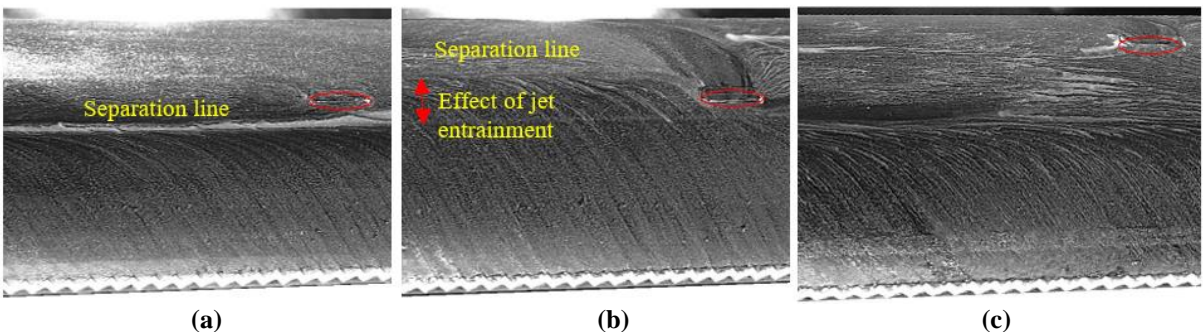

**Figure 14. Surface oil-flow patterns. a) $C_\mu= 0\%$; b) $\zeta_o=120°$ $C_\mu= 2.73\%$, right $\zeta_o=150°$ $C_\mu= 2.73\%$.**

Close-up photographs of the oil flow taken from another vantage point (fig. 15) looking at the lee of the cylinder offer a potential explanation for the sensitivity of the flow response to $\zeta_o$ location. At $\zeta_o=120°$ (fig. 15a) the emerging jet interacts with a relatively high speed spanwise flow coming out from the root, and gets overturned creating a swirling jet whose foot print is marked by 'S' shape streakline. Such counter-rotating swirl, at an appropriate $\zeta_o$ location could entrain the vortex fluid, as evidenced from the PIV data for $\zeta_o=120°$ (fig. 16b), leading to the

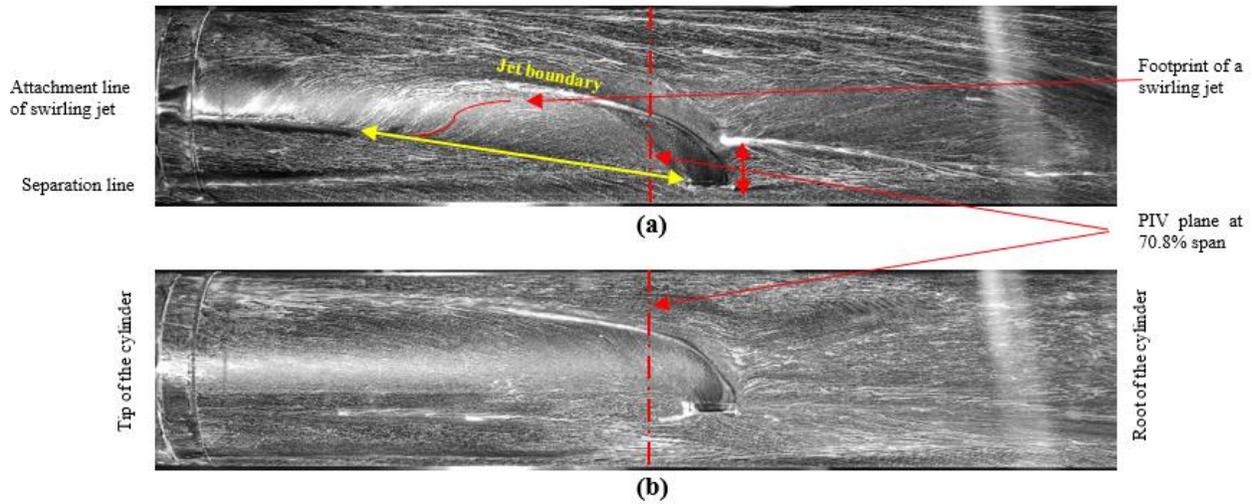

**Figure 15. Comparing two foot-prints of sweeping jets. a) jet emanating from $\zeta_o=120°$; b) jet emanating from $\zeta_o=150°$.**

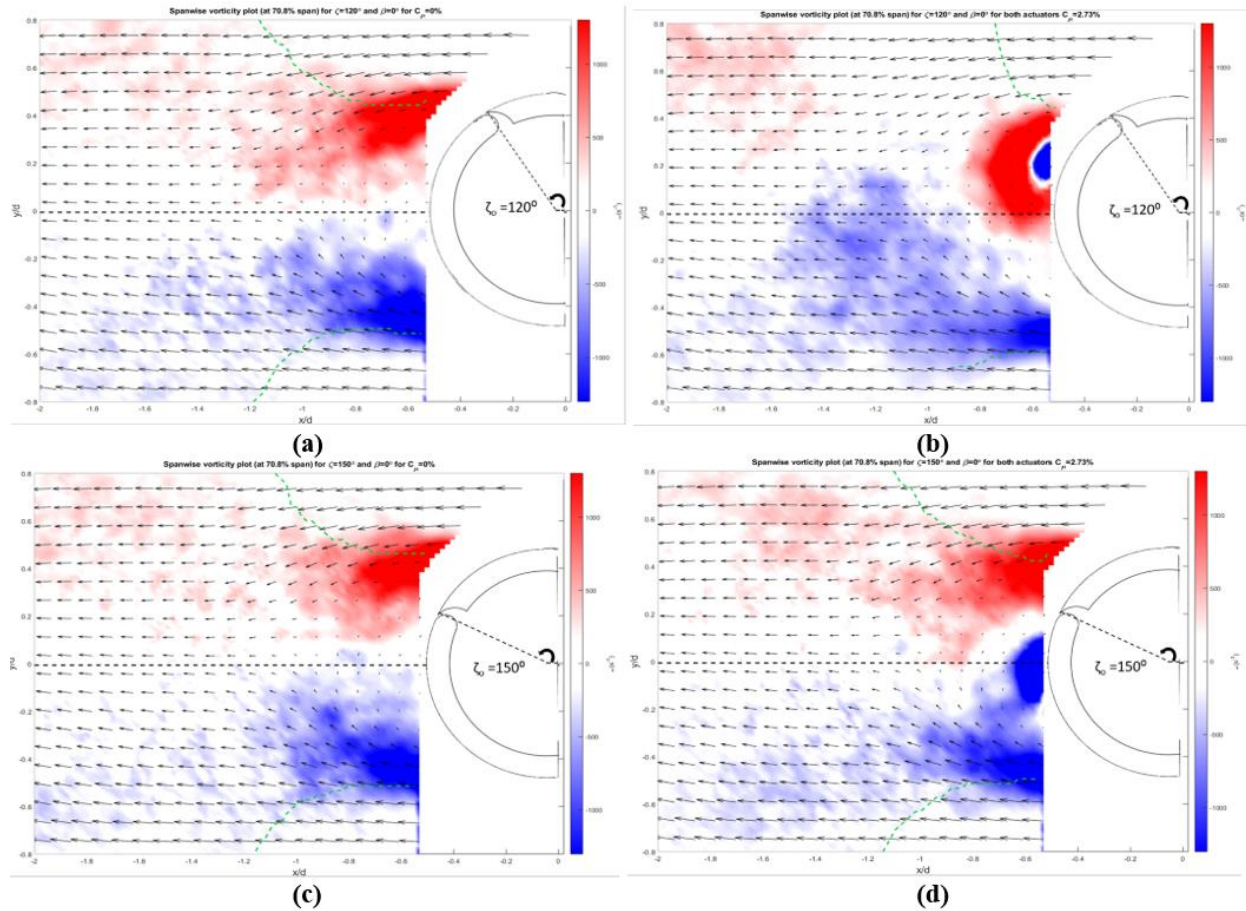

**Figure 16. Spanwise vorticity contour at 70.8% span at $\zeta_o=120°$ for a) $C_\mu= 0\%$, b) $C_\mu= 2.73\%$; and at $\zeta_o=150°$ for c) $C_\mu= 0\%$, d) $C_\mu= 2.73\%$; with the normalized wake half-width highlighted by green broken lines.**

reattachment of the vortical fluid. Such boundedness of the top vortical flow, entrains the high speed freestream fluid from the top-half of the cylinder, and pulls it downstream in the azimuthal direction thus delaying the occurence of separation. This was supported by the investigation of normalized velocity in the azimuth direction at various *x/d*

American Institute of Aeronautics and Astronautics

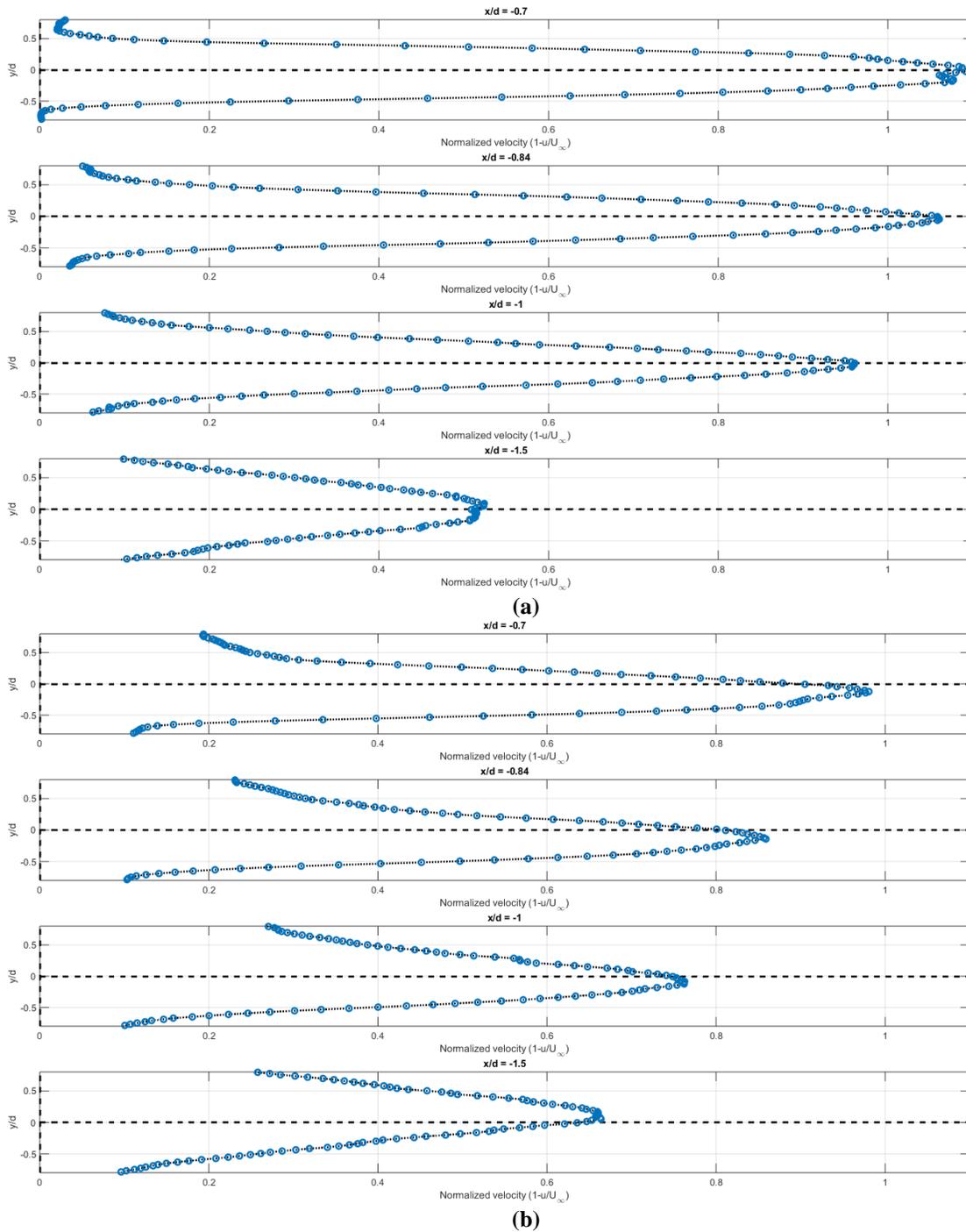

Figure 17. Normalized velocity in the azimuthal direction at various *x/d* positions, for $\zeta_o=120°$ a) $C_\mu= 0\%$, b) $C_\mu= 2.73\%$

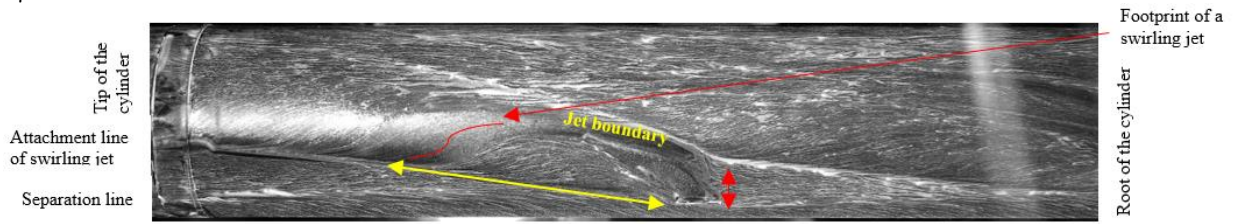

Figure 18. Footprints of the jet and the surrounding flow at $C_\mu= 0.95\%$ to emphasis similarity with $C_\mu= 2.73\%$
10
American Institute of Aeronautics and Astronautics

locations (fig. 17), where the normalized velocity was defined as $1 - u/U_{inf}$, $u$ indicates the azimuth velocity obtained from PIV experiments. Figure 17a indicates a good symmetry of the $C_\mu$= 0% flow at $\zeta_o$= 120º, as supported by $C_{LN}$= 0 for this value of $\zeta_o$ and $\beta$, in fig. 7. With actuation of $C_\mu$= 2.73% (fig.17b), the velocity profiles become highly asymmetric for all the $x/d$ locations examined. The magnitudes of the normalized velocity for $y/d$=0 drops from ~[1.5, 1.4, 0.96, 0.5] for $C_\mu$= 0% to ~[0.98, 0.85, 0.76, 0.68] for $C_\mu$= 2.73%, at subsequent $x/d$ locations, supporting the entrainment of the high speed flow principle. Wake half-width $y_w$ (normalized by $d$), another parameter considered for comparsion, was defined as the distance between the $y/d$ location of maximum normalized velocity in the wake, to the $y/d$ location of 50% of that maxima. For $C_\mu$= 0%, at both $\zeta_o$= 120º and $\zeta_o$= 150º had symmetric wake half-widths, represented by green broken lines in fig. 16a and 16c respectively. For $C_\mu$= 2.73% at $\zeta_o$= 120º, the wake half-widths were no longer symmetric with $y/d$=0 (fig. 16b). The difference in the wake half-width ($y_{+1/2} - y_{-1/2}$) is plotted in fig. 19 to emphasize the asymmetry generated due to actuation. For both $\zeta_o$= 120º and $\zeta_o$= 150º, at $C_\mu$= 0%, ($y_{+1/2} - y_{-1/2}$)≈0. For $\zeta_o$= 120º (fig. 19a), at various configurations of $C_\mu$, the asymmetry in the wake is highlighted by non-zero ($y_{+1/2} - y_{-1/2}$). This asymmetry in the wake half-widths (figs. 16b and 19a) coupled with the asymmetry in the azimuth normalized velocity profiles (fig. 17b) as compared with the $C_\mu$= 0% (figs. 16a and 17a), results in a non-zero $\Delta C_{LN}$.

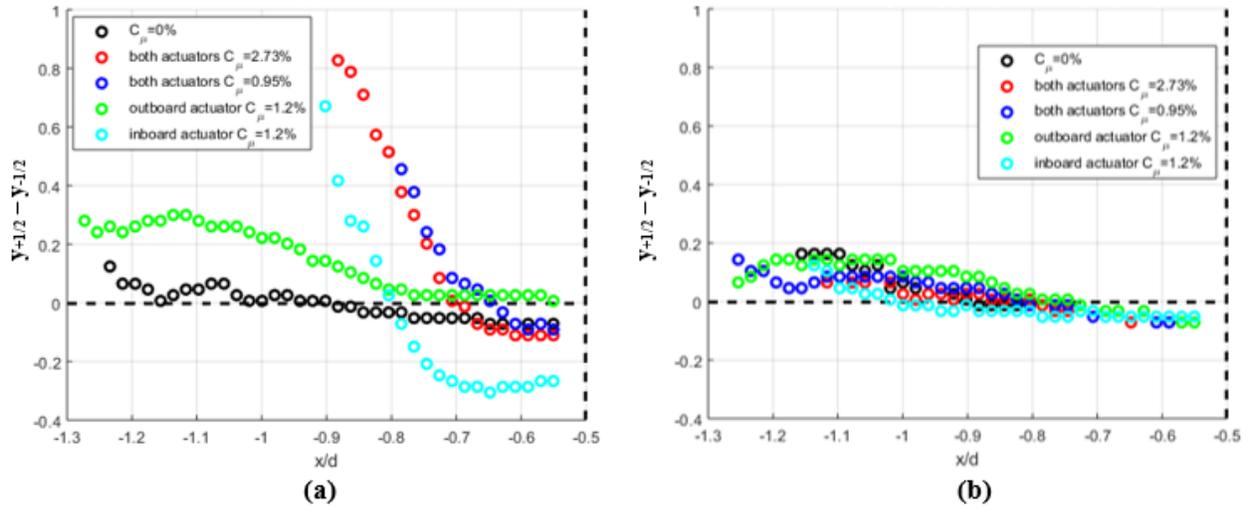

**Figure 19. Comparison of difference of the wake half-width ($y_{+1/2} - y_{-1/2}$) at $\beta$= 0º, various $C_\mu$ configurations, for a) $\zeta_o$=120º; b) $\zeta_o$=150º, to emphasize the significance of wake asymmetry in generating $\Delta C_{LN}$**

When the jet emerges at $\zeta_o$=150º where the spanwise flow is very sluggish, the jet penetrates farther along the azimuth with not so significant swirling strength (fig. 15b). Due to the actuator's azimuthal location and lower strength, it was observed to not be able to entrain the counter-rotating vortex fluid (fig. 16b), thus not affecting the location of separation. The marked wake half-width for $C_\mu$= 2.73% (fig. 16d) is almost symmetric, and ($y_{+1/2} - y_{-1/2}$)≈0 (fig. 19b); additionally, the symmetry of the azimuth normalized velocity was not significantly affected by the actuation, resulting in a negligible $\Delta C_{LN}$. This might be the reason for the success or failure of the actuators in modifying the baseline flow field.

This observation was further supported by investigating the magnitude of spanwise vorticity at all $x/d$ locations. For $C_\mu$= 0%, at $\zeta_o$= 120º and $\zeta_o$= 150º, the counter-rotating vortices were observed to be represented by rigid-body rotation, with its maxima at $y/d$≈ 0.4 and -0.45 (for $x/d$=0.6), and minima at $y/d$≈ 0 while maintaining symmetry. With actuation, for $C_\mu$= 2.73%, at $\zeta_o$= 120º, the location of the maxima were displaced to $y/d$≈ 0.3 and -0.55 respectively (for $x/d$=0.6), with the maximum of the swirling jet at $y/d$≈ 0.2. This 'intentional' asymmetry achieved, with respect to the geometrical axis of the cylinder, further supports the generation of yawing moment. While for $C_\mu$= 2.73%, at $\zeta_o$= 150º, no displacement of the maxima location was observed, thus maintaining the symmetry of the counter-rotating vortices.

By lowering the $C_\mu$ from 2.73% to 0.95% while maintaining all other parameters affecting the flow to be constant, the surface flow pattern did not change substantially nor did the yawing moment. In fact all that one needs to do is to rescale the ordinate of fig. 13 where $\Delta C_{LNmax}$= 0.92 for the various combinations of $\beta$ & $\zeta_o$ in fig. 13 will decrease to $\Delta C_{LNmax}$= 0.64. The oil flow pictures of fig. 15a and 18 suggest the following topology resulting from the Jet – Boundary Layer interaction and its similarity regardless of variation of $C_\mu$. The upstream jet boundary leaving the right hand corner of the nozzle turns over and seem to reach the surface at the end of the yellow double-arrow. The second surface streamline leaving the left nozzle corner meets the first one at that point and the two seem to form an attachment line that marks the left border of the jet (fig. 18). The distance required for the two surface streamlines, marking the jet boundaries to meet (i.e. the presumed first turnover of the jet), is marked by the length of the yellow arrow. It is affected by $C_\mu$, as is the width of footprint of the swirling jet and the distance between the separation and the reattachment line (see left hand side of fig. 15a and 18).

Since $\Delta C_{LN}$ was close to its maximum value for $\beta = -10°$ & $\zeta_o = 150°$, $\beta = 0°$ & $\zeta_o = 120°$ and for $\beta = 10°$ & $\zeta_o = 90°$ for the same input and location of both AFC actuators, it was instructive to compare the oil flow of these three cases (see the black arrows in fig. 13). Such a comparison indicates that substantial differences in the resulting flow pattern near the outboard actuator exist. These differences were attributed to the asymmetry generated by $\beta \neq 0°$ and the different azimuthal location of the actuators (marked in red in fig. 20). However, the sole salient feature that characterizes the three pictures is the footprint of the swirling jet (fig. 20). The generation of the swirl should depend on the location of the sweeping jet and its orientation relative to the oncoming flow and this could explain the differences in the receptivity encountered by the various wing platforms when sweeping jet actuation was applied to them.

It was interesting to note that when two actuators are used between $-10° < \beta < +10°$ one may attain almost identical $\Delta C_{LNmax}$ by rotating the cylinder between $135° < \zeta_o < 90°$ suggesting that it should be possible to yaw the cylinder to either side by rotating two actuators between these $\zeta_o$ limits. When a single jet was used this was no longer the case. For the outer actuator an increase in $\beta$ results in a lower $\Delta C_{LNmax}$ for a given $C_\mu$ input, in spite of the fact that $\zeta_o$ was decreased from 150° to 75° thus attaining its local $\Delta C_{LNmax}$ within the range of $\beta$ of interest (fig. 21a). The increase in $\beta$ was responsible for moving the separation line upstream. Thus for a given $\zeta_o=120°$ the $\Delta C_{LN}$ may be substantial or deteriorate by changing $\beta$ (see the 3 purple dots on the $\zeta_o=120°$ curve in fig. 21a). For $\beta= -10°$ the sweeping jet interacted with attached flow but close enough to the separation line to be effective (fig. 21b, top). One may note that the slope $(d\Delta C_{LN}/d\beta) > 0$ suggesting that $\Delta C_{LNmax}$ was not reached. At $\beta= 0°$ it was already in the separated zone but very close to the separation line in absence of blowing, but blowing moved the separation line upstream on the inboard side of the nozzle while the active jet pulled the separation line back to the nozzle and beyond it, on the outboard side of the nozzle and so $(d\Delta C_{LN}/d\beta) < 0$. For $\beta= +10°$ the nozzle was so far azimuthally downstream of the separation line that blowing at $C_\mu=1.2\%$ did not affect its location (fig. 21b, bottom). Consequently, the location of the sweeping jet in relation to the separation line was of outmost significance when separation line was determined by stagnation of the azimuthal component of the flow at the surface.

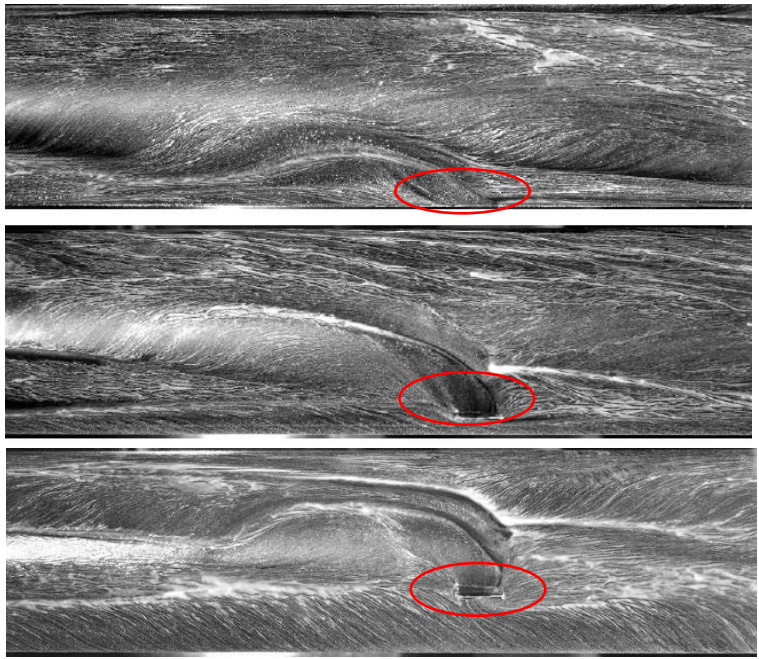

**Figure 20. Footprints of the jet and the surrounding flow for $\beta =10°$ & $\zeta_o=90°$ (top), for $\beta = 0°$ & $\zeta_o=120°$ (center), $\beta = -10°$ & $\zeta_o=150°$ (bottom), all generate the same $\Delta C_{LN}$ at $C_\mu=2.73\%$ when two actuators were utilized**

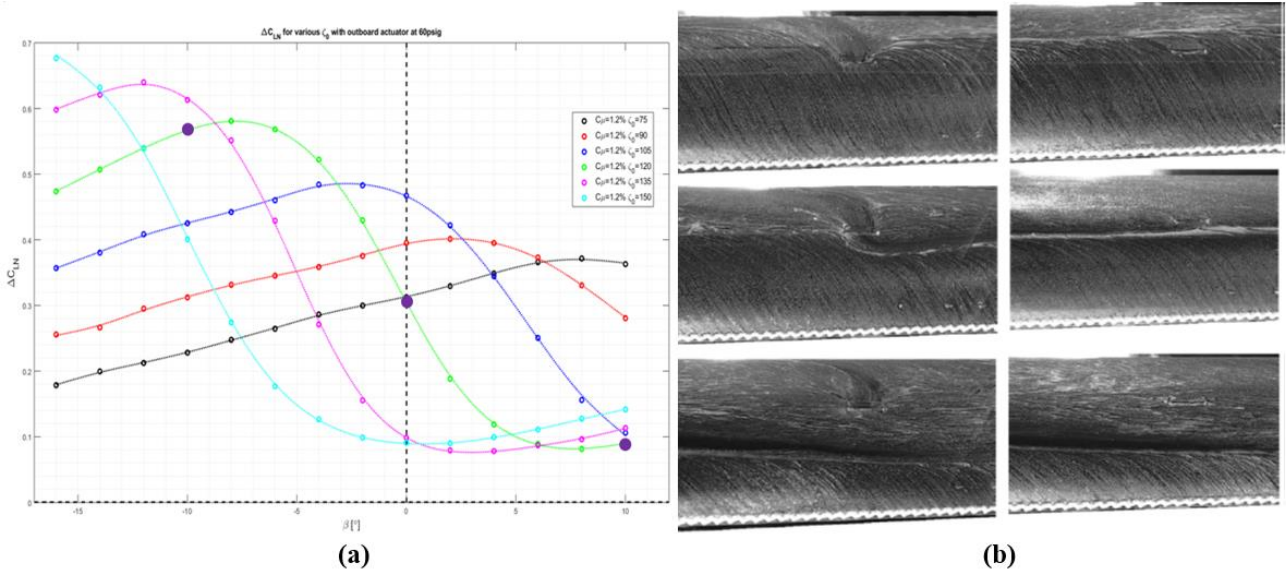

**Figure 21. a)** $\Delta C_{LN}$ **variation with** $\beta$ **&** $\zeta_o$ **for outboard actuator at** $C_\mu$=1.2%**. Purple dots on** $\zeta_o$=120º **correspond to b) the 6 pictures at** $\beta$= -10º **(top); 0º (center); +10º (bottom). Right:** $C_\mu$= 0%**; Left:** $C_\mu$= 2.73%**.**

When only the inner actuator jet was used, an increase in $\beta$ results in a concomitant increase in $\Delta C_{LNmax}$ for a given $C_\mu$=1.2% when $\zeta_o$ was decreased from 135º to 75º (fig. 22). For $\zeta_o$=150º $(d\Delta C_{LN}/d\beta) < 0$ for $\beta$> -14º. The increase in $\beta$ was still responsible for moving the baseline separation line azimuthally upstream (to smaller $\zeta$), as it was when the outboard actuator was considered. At $\zeta_o$= 90º, $\Delta C_{LN}$ undergoes a large increase, reaching $\Delta C_{LNmax}$ at $\beta$= +10º and its slope $(d\Delta C_{LN}/d\beta) > 0$ throughout the region of interest. At this yaw angle the baseline separation line approximately coincided with the inboard nozzle location and therefore the flow was receptive to AFC. At $\beta$= -10º the nozzle of the inboard actuator was embedded deeply in the attached flow region where the direction of the flow was more aligned with the sweeping jet axis than being orthogonal to it. Consequently AFC was ineffective (fig. 20). The picture inserted in fig. 22 corresponds to the surface flow photographed at $\beta$= +10º when the inboard actuator was active at $C_\mu$= 1.2% (cannot be seen in the picture). Once again the flow pattern suggests that a swirling jet was formed spanwise downstream of the actuator.

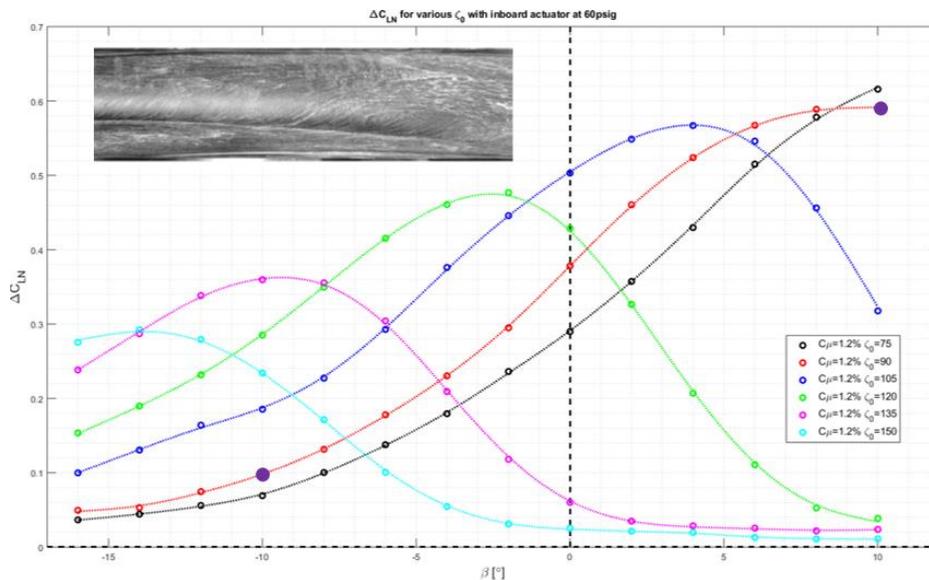

**Figure 22.** $\Delta C_{LN}$ **variation with** $\beta$ **&** $\zeta_o$ **for the inboard actuator at** $C_\mu$=1.2%**. Note purple dots on** $\zeta_o$=90° **curve.**

One may cross plot the results shown in figs. 13, 21 & 22 for 3 yaw angles $\beta$= -10º; 0º & +10º as shown in fig. 23. When two sweeping jets were used $\Delta C_{LNmax} \approx 0.9$ regardless of $\beta$. However, while for $\beta$= -10º $\Delta C_{LNmax}$ corresponded to $\zeta_o$= 135º for $\beta$=0º it did so at $\zeta_o$=115º and for $\beta$=+10º the best $\zeta_o$=90º. Furthermore since at $\beta \neq 0$º the natural mean separation line was no longer parallel to the cylinder axis but varied along the span, the outboard actuator was more effective at $\beta$< 0º, while the inboard actuator was more effective at $\beta$> 0º; and at $\beta$= 0º there is no recognizable difference between these two spanwise locations (fig. 23b). The difference was attributed to the azimuthal distance between the local separation line and the location of the actuator. One may note that if the actuator was located in the attached flow zone, $(\partial \Delta C_{LN}/\partial \beta)_\zeta >0$ and $(\partial \Delta C_{LN}/\partial \zeta)_\beta >0$ and they both change sign when the actuation is located downstream of separation location. The magnitude of the positive slope $(\partial \Delta C_{LN}/\partial \beta)_\zeta >0$ in figs. 13 & 22 in particular was lower than the negative one suggesting that attached flow is less sensitive to the specific location of the actuator relative to the separation line than the separated flow is. The results of fig. 13 suggest that two actuators dividing the span into three equal segments that can be rotated around the axis of the cylinder (i.e. in the azimuthal direction) can maintain a prescribed yawing moment regardless of the sign and magnitude of the maximum $\beta$ tested to date (i.e. -15º $< \beta < $ 15º).

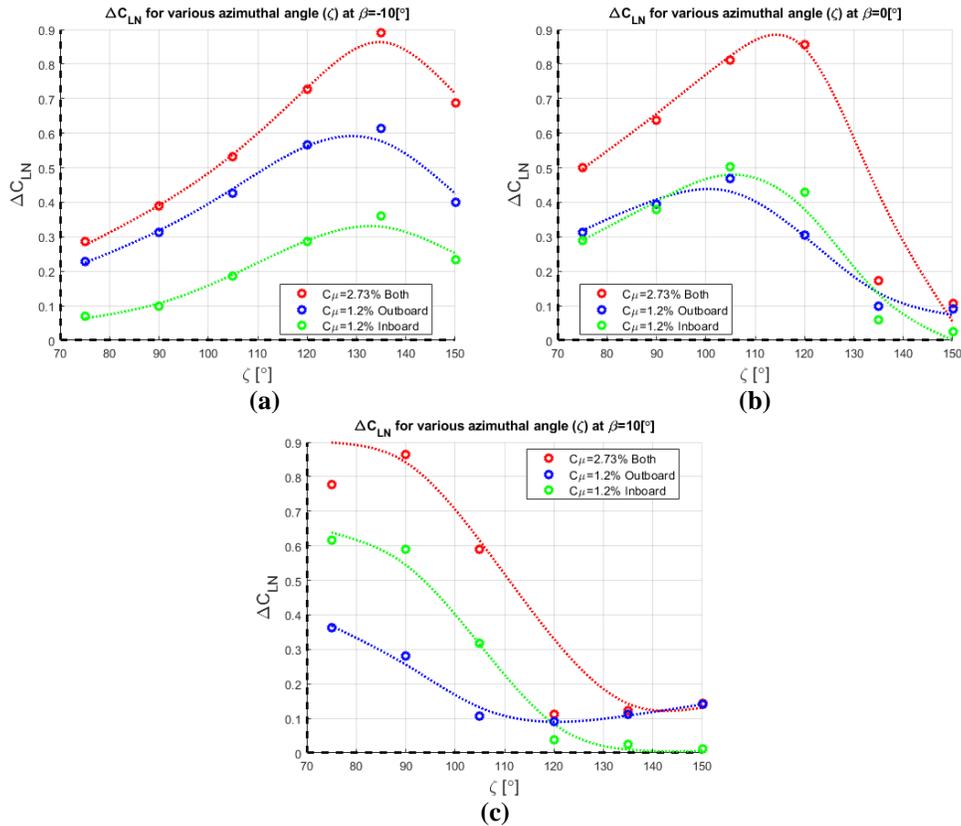

**Figure 23. $\Delta C_{LN}$ Vs. $\zeta_o$ to investigate individual actuator performance at a) $\beta$= -10º, b) $\beta$=0º, and c) $\beta$=10º**

## V. Conclusion

The effects of active flow control being applied to a swept-back cylinder was investigated using a five-component force balance, oil flow visualization and some basic PIV. The baseline flow became Reynolds number independent provided there was a trip strip along both the circumference at the root of the cylinder and additional two trip strips along the span. Under these conditions the separation line in the absence of yaw was parallel to the cylinder axis (i.e. for $\beta = 0$º $\zeta_{sep} \approx 115$º). It appears that a single sweeping jet whose axis is orthogonal to the surface separation line and whose nozzle is located on or near the separation line is most effective in controlling the flow and providing a side force and a yawing moment of substance. Two jets dividing the span into three equal segments do better and the result is independent of the sign of the yaw angle within the range of $\beta$= ±15° that were tested. At $\beta = 0$º and fixed location of the actuators (e.g. $\zeta$=90°) the measured yawing moment, $\Delta C_{LN}$ was proportional to $\sqrt{C_\mu}$, suggesting that jet entrainment might be partially responsible for generating side force. However, surface flow suggested that the jet swirls due to its interaction with the spanwise flow and it turns downstream into a vortex that is approximately aligned with the spanwise flow. The result was sensitive to the location of the actuator relative to the separation line. An actuator located downstream of the separation line could become totally ineffective within a short distance. Consequently there is a close coupling between a single vortex (swirling jet) originated by the sweeping jet and the yawing moment generated by it. This proposition has to be further investigated by mapping the flow at its entirety. It appears that sufficiently large yawing moment coefficients can be generated to be used in the control of a refueling boom, thus avoiding the 'H' shape control surfaces that are currently used on the newer Air Force tankers.